\documentclass{article}

\usepackage{amstext}
\usepackage{amsfonts}
\usepackage{amssymb}
\usepackage{amsmath}
\usepackage{graphics}
\usepackage{svg}
\usepackage{url}
\usepackage{color}

\begin{document}

\title{A-Phase classification using convolutional neural networks}
\author{Edgar R. Arce-Santana\thanks{E-mail: arce@fciencias.uaslp.mx} \and Alfonso Alba \and Martin O. Mendez \and Valdemar Arce-Guevara  \\ \\ 
Laboratorio Nacional Centro de Investigaci\'on en Imagenolog\'ia \\ e Instrumentaci\'on M\'edica, \\ 
Facultad de Ciencias \& CICSaB, \\
Universidad Aut\'onoma de San Luis Potos\'i, M\'exico \\ 
Tel.: +52-444-8262300 x 5623 \\
}
              
\date{July 22, 2019}

\maketitle

\begin{abstract}
A series of short events, called A-phases, can be observed in the human electroencephalogram during NREM sleep. These events can be classified in three groups (A1, A2 and A3) according to their spectral contents, and are thought to play a role in the transitions between the different sleep stages. A-phase detection and classification is usually performed manually by a trained expert, but it is a tedious and time-consuming task. In the past two decades, various researchers have designed algorithms to automatically detect and classify the A-phases with varying degrees of success, but the problem remains open. In this paper, a different approach is proposed: instead of attempting to design a general classifier for all subjects, we propose to train ad-hoc classifiers for each subject using as little data as possible, in order to drastically reduce the amount of time required from the expert. The proposed classifiers are based on deep convolutional neural networks using the log-spectrogram of the EEG signal as input data. Results are encouraging, achieving average accuracies of 80.31\% when discriminating between A-phases and non A-phases, and 71.87\% when classifying among A-phase sub-types, with only 25\% of the total A-phases used for training. When additional expert-validated data is considered, the sub-type classification accuracy increases to 78.92\%. These results show that a semi-automatic annotation system with assistance from an expert could provide a better alternative to fully automatic classifiers.
\end{abstract}

\section{Introduction}

During the past decades, several studies have dedicated efforts to analyze and understand the electrical information of the brain during sleep. This has permitted important advances in the interpretation of normal and pathologic sleep. For example, the effects of a fragmented sleep have been correlated to pathologies such as metabolic syndrome \cite{wolk2007sleep}, irritability, lack of concentration and traffic accidents \cite{dinges1997cumulative,altevogt2006sleep}. 

Sleep evaluation is carried out by a clinical procedure called Polysomnography (PSG), which involves the acquisition of electrophysiological signals: electroencephalogram (EEG), electrooculogram and electromyogram. Thereafter, these signals are used to segment the sleep time in stages (wake, 1-4 and REM) based on the frequency content and temporal evolution of those signals. Furthermore, when detailed information related to a specific pathology (i.e. sleep apnea) is needed, other signals such as electrocardiogram, airflow and pulse oxymetry could be also recorded \cite{iber2007aasm}.
 In most cases, the sleep process has been mainly characterized by the sleep stages. However, nowadays there exists evidence of a  oscillatory brain pattern composed by short events that seem to be correlated with the dynamic of the sleep stages, as they play a role in maintaining a sleep stage as well as generating the sleep stage transitions \cite{terzano2002}. These short events, lasting between 2s and 60s, are observed as oscillations that disrupt the basal EEG rhythms of the sleep stages and are called A-phases. Each A-phase presents particular spectral characteristics and duration, which allow a subdivision in three groups:
\begin{itemize}
\item A1-phase. Characterized by bursts and K-complexes of Delta waves (0.5 Hz - 4 Hz).
\item A2-phase. Presents rapid EEG waves (Alpha (8 Hz - 12 Hz) and Beta (12 Hz - 30 Hz)) that cover between 20\% and 50\% of the A-phase duration and the Delta waves in the rest of the event duration.
\item A3-phase. Characterized by Alpha and Beta waves, which cover more than 50\% of the A-phase duration. 
\end{itemize}

In addition, A-phases are organized following some rules giving place to the oscillatory pattern named the Cyclic Alternating Pattern (CAP) \cite{terzano2002}. In recent studies, a clinical index based on the percentage of the sleep time in CAP (CAP index) has showed a high correlation with the sleep quality \cite{terzano1993clinical}. This shows the importance of CAP evaluation in clinics. However, even if CAP index seems to add valuable information about the sleep process with respect to the classical procedure, the detection and classification of the A-phases is a tedious and long visual procedure, requiring many months of training before becoming an expert clinician in the field. In addition, the A-phase annotation procedure suffers from large subjectivity and as a consequence there exists a high inter-scorer variability \cite{ferri2005}.

To alleviate this problem, some studies have analyzed the A-phases and their surroundings using various techniques, including spectral decomposition and complexity measures, from which one can obtain interesting characteristics that support the visual annotations \cite{ferri2005all,de2004quantitative}. Other studies have developed computational methods to automatically detect  the A-phases based on modern signal processing techniques such as wavelet analysis, and machine learning procedures, e.g. Neural Networks and Support Vector Machines. Their results reach levels of A-phase detection up to 78\%, with or without distinction of the A-phase type during the classification process \cite{navona2002,mariani2011,mariani2012}.

Until recently, A-phase detection sequentially follows  the steps of feature extraction,
selection, and classification, typically applied in data science. However,  new developments
in the machine learning field , in particular regarding deep learning, have led to  models
with the capability to carry out these steps together in a more efficient way \cite{goodfellowdeep}.
Deep learning  has achieved  an important success in many application areas such
as image recognition, sound processing and  natural language processing. Specifically, there has been a wide adoption of deep learning approaches to evaluate, characterize and classify biomedical signals (EEG, ECG, EMG,
 EOG, MRI and CT) \cite{faust2018deep,litjens2017survey,shen2017deep}. We can find applications of deep learning methods in many challenging problems such as automatic ECG evaluations in heart diseases \cite{yildirim2018novel,acharya2017deep,yildirim2018arrhythmia} and  pathologic event detection  at EEG \cite{oh2018deep,acharya2018deep,yildirim2018deep,antoniades2018deep} . There are also a few studies  with deep learning models  for the sleep stage classification. For instance, Supratak et al. \cite{supratak2017deepsleepnet} presented a method using the combination of convolutional neural network (CNN) and bidirectional long short- term
memory (BLSTM), Tsinalis et al. \cite{tsinalis2016automatic} applied a CNN  on single EEG channel,  Tripathy and Acharya \cite{tripathy2018use}  
used  RR-time series and EEG signals, Chambon et al. \cite{chambon2018deep}
applied  two-dimensional (2D) CNN  model  and Michielli et al. \cite{michielli2019cascaded} used a cascaded LSTM architecture. However, the application of deep learning techniques for the detection and classification of A-phases has not yet been fully explored.

For this reason, this study attempts to evaluate the feasibility of classifying the EEG acquired during sleep in a two-fold task: 1) discriminate between A-phase and not A-phase (N-phase) segments, and 2)  classify  the A-phase segments as A1-phase, A2-phase or A3-phase. The time-frequency decomposition (spectrogram)  for each 4s EEG segment is used as input to a CNN  architecture that is trained for each subject using a reduced amount of expert annotations, and possibly refined under expert supervision. Unlike previous works, where EEG segments of 1 second are often used for A-phase classification, we decided instead to use 4-second segments since we are assuming that the potential onset times of A-phases are known in advance. The detection of the onset times is a problem that will be treated in a future publication.

\section{Previous works}

In the past 20 years, various works have been published with the aim of producing an automatic or semi-automatic scoring of the cycling alternating pattern from EEG signals. This is a very complex task, which is why many of these works focus on different sub-tasks; for example, some works focus on simply detecting phasic events (A-phases), while others focus on characterizing and classifying the different types of events (e.g., A1, A2 and A3-phases). In most cases, the accuracy of the results lies within the expected range of inter-scorer agreement, and in this sense, they can be considered successful. On the other hand, only a few proposals have been able to produce outstanding results, which is probably due to most classifiers having problems to generalize their models to data from new subjects.

One of the early works, by Rosa et al. in 1999 \cite{rosa1999}, proposes to first model the EEG signal using a bank of parallel bandpass filters tuned to the classic EEG rhythms, whose gains are estimated by maximum likelihood \cite{rosa1991}. In this way, a vector of gains for delta, theta, alpha and beta activity is obtained for each 1s segment of the EEG signal. A matched filter is then applied to the linear combination of the estimated gains in order to detect A-phases, using pulse waves with increasing lengths as templates. This results in a binary signal which is then fed to a state machine that outputs a post-processed sequence ensuring that the duration of A and B-phases is between 2 and 60 s. No  distinction is made between A1, A2 and A3 phases. The authors report an average accuracy of 89.8\% from tests with four subjects with no sleep pathologies.

Navona et al. proposed using spectral EEG features obtained by averaging the amplitude of the output of a bank of bandpass filters (also tuned at the classic EEG rhythms) within segments of two different lengths: 2s (short-window) and 64s (long-window), computed every 0.5s. A descriptor for each band is then computed as the relative difference between the short-window and long-window averages. The idea is to characterize how much the local activity differs from the surrounding background activity. A simple heuristic based on thresholds is applied to detect A-phases from the spectral descriptors. Further heuristics are applied to classify each A-phase as A1, A2 or A3. Results from 10 subjects (where specific NREM segments were chosen from different sleep-cycles) showed an accuracy of 77\% for A-phase detection and 79\% for A-phase classification \cite{navona2002}. An updated version of this method was later proposed by Barcaro et al., reporting accuracies of 83.5\% for the A-phase detection stage, and 73.7\% when A-phase classification was also taken into account \cite{barcaro2004}.

Among the first works that used trained classifiers instead of heuristics for A-phase detection and classification, are those by Mariani et al. The authors computed several features from the EEG signals, including spectral descriptors similar to those suggested by Navona et al., and used these data to train different types of classifiers, such as linear and quadratic discriminants (LDA/QDA), feed-forward neural networks, support vector machines (SVM) and adapting boosting (AdaBoost). Each 1s segment of the EEG signal was thus classified as belonging to an A-phase or not. The best results were obtained with a linear discriminant and a feed-forward neural network, yielding accuracies of 84.9\% and 81.5\%, respectively \cite{mariani2012}.

By 2012, the Physionet CAP Sleep database was released to the public, allowing further studies to be conducted by research groups from many different institutions. Mendez et al. have explored the discriminating power of several EEG features, including spectral, statistical, entropy and complexity features, using simple classifiers such as k-Nearest Neighbors (kNN). These results show high accuracy values (89\% to 94\%, in average) when discriminating between the B-phase and A-phase around the onset of an A-phase, and competitive accuracies (80\% to 87\%) when discriminating between the A-phase and B-phase around the end of the A-phase, suggesting that a scoring approach based on change-point detection could produce good results \cite{mendez2014,mendez2016}. More recent works have explored more combinations of features and classifiers \cite{karimzadeh2015,mendonca2018b}, but accuracy results have not been improved and remain between 70\% and 80\%.

For most of the works mentioned above, the focus lies in the detection of A-phases; while only a few publications deal with the classification of A-phases according to their sub-type (A1, A2 and A3). Among the latter, it is worth noting the works by Barcaro et al. \cite{navona2002,barcaro2004} and by Mendez et al. \cite{mendez2014}, which have already been discussed. Barcaro proposes an heuristic process to classify the A-phases, whereas Mendez studies the discriminative power of several features for A-phase classification (obtaining an accuracy of 82.23\% for the training data). A similar work by Machado et al. uses a set of 55 features to classify the A-phases between two groups: A1 and A2/A3, using different types of classifiers, including quadratic discriminant, k-Nearest Neighbors and Support Vector Machines (SVM). The best result is obtained with an SVM, achieving an accuracy of 71\% \cite{machado2016,machado2018} for the classification of A-phases subtypes and 76\% for discriminating between A-phases and B-phases. This suggests that distinguishing among the sub-types of A-phases could be a harder problem than distinguishing between A-phases and B-phases. Finally, Mostafa et al. employed a deep neural network to classify 2s segments as A-phases or non A-phases, achieving an accuracy of 67\% \cite{mostafa2018}. This approach is particularly interesting since it is not based on computing a large set of features; instead, the EEG signal is fed into the network, and the first layers of the network learn how to encode the signal for the classification stages. We believe similar approaches should be further explored.

Based on the results of the works discussed above, automatic classification of A-phases seems to be a very difficult problem. One of the factors that make this problem so difficult is the large variability between subjects and acquisition protocols; indeed, the Physionet CAP database contains registers using different EEG traces and different sampling frequencies. On the other hand, given that the CAP inter-scorer agreement lies around 70\% \cite{ferri2005}, there is some uncertainty about the expert annotations used to train the automatic classifiers. For these reasons, designing a fully automatic classifier that correctly classifies data from new subjects, is a very complex task. A more practical approach could be to design an ad-hoc classifier for each subject, based on a small set of annotations for that same subject. In this way, an expert scorer could annotate only a few representative A-phases, and then let the system detect and classify the rest, with some degree of confidence. While such a system would not be fully automatic, it could still save plenty of the expert's time.

\section{Methodology}

In this section, we describe data and used methodology to classify EEG segments as A-phases and non A-phases (N-phases), as well as A-phases according to their sub-type (A1, A2 and A3). In this approach a Deep-Learning strategy is used, specifically, we propose to train a Convolutional Neural Networks (CNN) as classifier. First, we define an architecture to distinguish between A-phases and N-phases; second, another network is trained to classify sub-types A-phases.

\subsection{Database}

Polysomnographic recordings of nine healthy subjects were used in this study. The recordings belong to four females and five males, with ages between 23 and 37 years old (mean = 31.66 $\pm$ 4.27). The dataset is freely available from the CAP Sleep Database of Physionet \cite{goldberger2000} and has a number of one-night polysomnographic recordings from both normal and pathologic subjects. These recordings were registered and annotated at the Sleep Disorders Center of the Ospedale Maggiore of Parma, Italy. Data includes information of at least three EEG channels (F3 or F4, C3 or C4 and O1 or O2), three electromyography signals,  two electrooculographic channels, respiratory signals and electrocardiogram. The recordings presented different sampling rates  between 256 and 512 Hz. However, all the recordings were resampled at 512 Hz using cubic spline interpolation.

The scoring for macrostructure and microstructure was performed by  expert neurophysiologists and this information is also available with the data. The macrostructure was annotated according to the R\&K rules \cite{iber2007aasm}, while CAP was annotated in agreement with Terzano reference atlas  \cite{terzano2002}.

\begin{figure*}[h!]
       \centering
        \includegraphics[width=12cm, height=10cm]{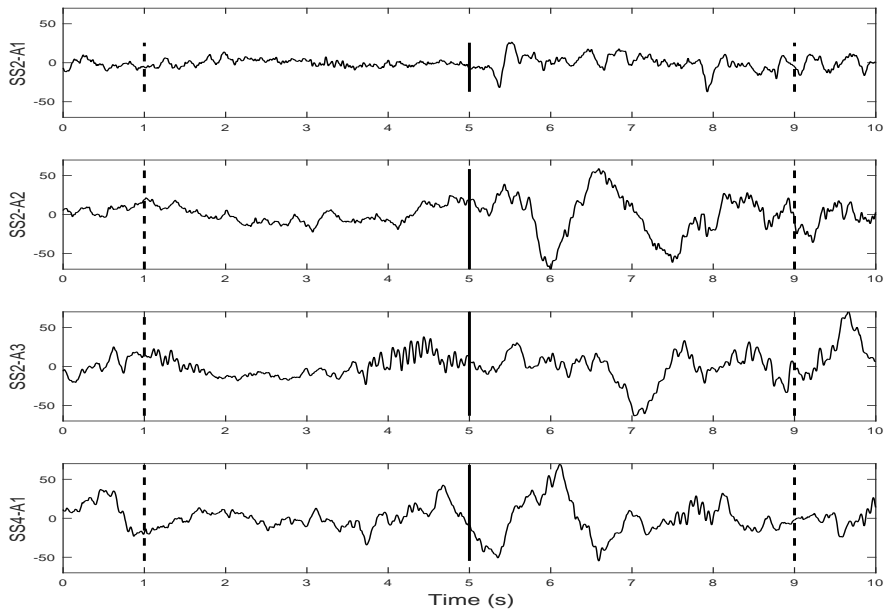}
        \caption {Example of A1-, A2- and A3-phases during sleep stage 2 (SS2) and 4 (SS4). The vertical black line indicates the onset of the A-phase and  the vertical dashed lines are placed four seconds before and after A-phase onset, delimiting the N-phase and A-phase segments. }
        \label{figAphase}
    \end{figure*}

From each A-phase two segments were extracted: a) four seconds before the A-phase onset (N-phase) and  b) four seconds after the A-phase onset (A-phase). Fig.  \ref{figAphase} shows examples of A-phases at sleep stage 2 (SS2) and 4 (SS4). The vertical black line is the A-phase onset. The segment from the left dashed line until the onset corresponds to the N-phase, whereas the segment from  the onset until the right dashed line corresponds to A-phase segments used in this study.  A total of 2887 N-phases and 3690 A-phases were obtained from the recordings. There is a smaller number of N-phases since we discarded the N-phases that overlapped with a previous A-phase (recall that the distance between two consecutive A-phases can be as small as 2s). For each A-phase type we have 2373 segments of type A1, 680 of type A2¡ and 637 of type A3. A summary of the data per subject can be found in Table \ref{tabAphase}. 

\begin{table*}[h]
\caption{Number, total time and average duration of A-phases for each subject and each A-phase type (A1/A2/A3). Duration is shown as average $\pm$ standard deviation.}
\label{tabAphase}
\scriptsize
\vspace{0.01cm}
\begin{tabular}{|c|c|c|c|c|c|c|c|c|c|}
\hline
& \multicolumn{3}{|c|}{Number of A-phases} & \multicolumn{3}{|c|}{Total time in A-phase(s)} & \multicolumn{3}{|c|}{Duration A-phase(s)} \\
\hline
& \multicolumn{1}{c}{A1}& \multicolumn{1}{c}{A2} & A3 & \multicolumn{1}{c}{A1} & \multicolumn{1}{c}{A2}& A3 & \multicolumn{1}{c}{A1}& \multicolumn{1}{c}{A2} & A3\\
\hline
S1 & 363 & 94 & 80 & 2217  & 747 & 1135 & 6.11 $\pm$ 3.32 & 7.95 $\pm$ 4.48 & 14.19 $\pm$ 9.59\\
S2 & 186 & 72 & 94  & 1188 & 688  & 1239 & 6.39 $\pm$ 3.05 & 9.56 $\pm$ 4.94 & 13.18 $\pm$ 6.88\\
S3 & 141 & 106 & 108 &  656 & 631 & 1043 & 4.65 $\pm$ 2.17 & 5.95 $\pm$ 3.98 & 9.66 $\pm$ 8.04\\
S4 & 462 & 24 & 60 &  2863 & 328 & 784 & 6.20 $\pm$ 3.01 & 13.67 $\pm$ 8.33 & 13.07 $\pm$ 8.21\\
S5 & 164 & 34 & 61 &  1489 & 336 & 922 & 9.08 $\pm$ 3.53 & 9.88 $\pm$ 4.23 & 15.11 $\pm$ 10.00\\
S6 & 235 & 80 & 50 & 1724 & 583 & 796 & 7.34 $\pm$ 3.25 & 7.29 $\pm$ 5.12 & 15.92 $\pm$ 9.37\\
S7 & 303 & 115 & 94 & 1871 & 976 & 1414 & 6.17 $\pm$ 2.51 & 8.49 $\pm$ 3.60 & 15.04 $\pm$ 9.43\\
S8 & 307 & 99 & 42  & 1616 & 565 & 480 & 5.26 $\pm$ 2.75 & 5.71 $\pm$ 3.39 & 11.43 $\pm$ 10.65\\
S9 & 212 & 56 & 48 & 1036 &  377 & 678 & 4.89 $\pm$ 1.96 & 6.73 $\pm$ 3.69 & 14.12 $\pm$ 11.55\\
\hline
\end{tabular} 
\end{table*}

\subsection{Methods}
In this section, we describe the methods used  to classify A-phases and N-phases, as well as A-phases according to their sub-type (A1, A2 and A3). The classification procedure is carried out with a deep learning strategy; specifically, we propose to train an ad-hoc Convolutional Neural Network (CNN) for each subject. First, we define an architecture to distinguish between A-phases and N-phases; second, another network is trained to classify A-phase types.

\subsubsection{Preprocessing}
\begin{figure}[t]
\centering
\begin{minipage}[b]{0.45\linewidth}
  \centering
  \centerline{\includegraphics[width=\linewidth]{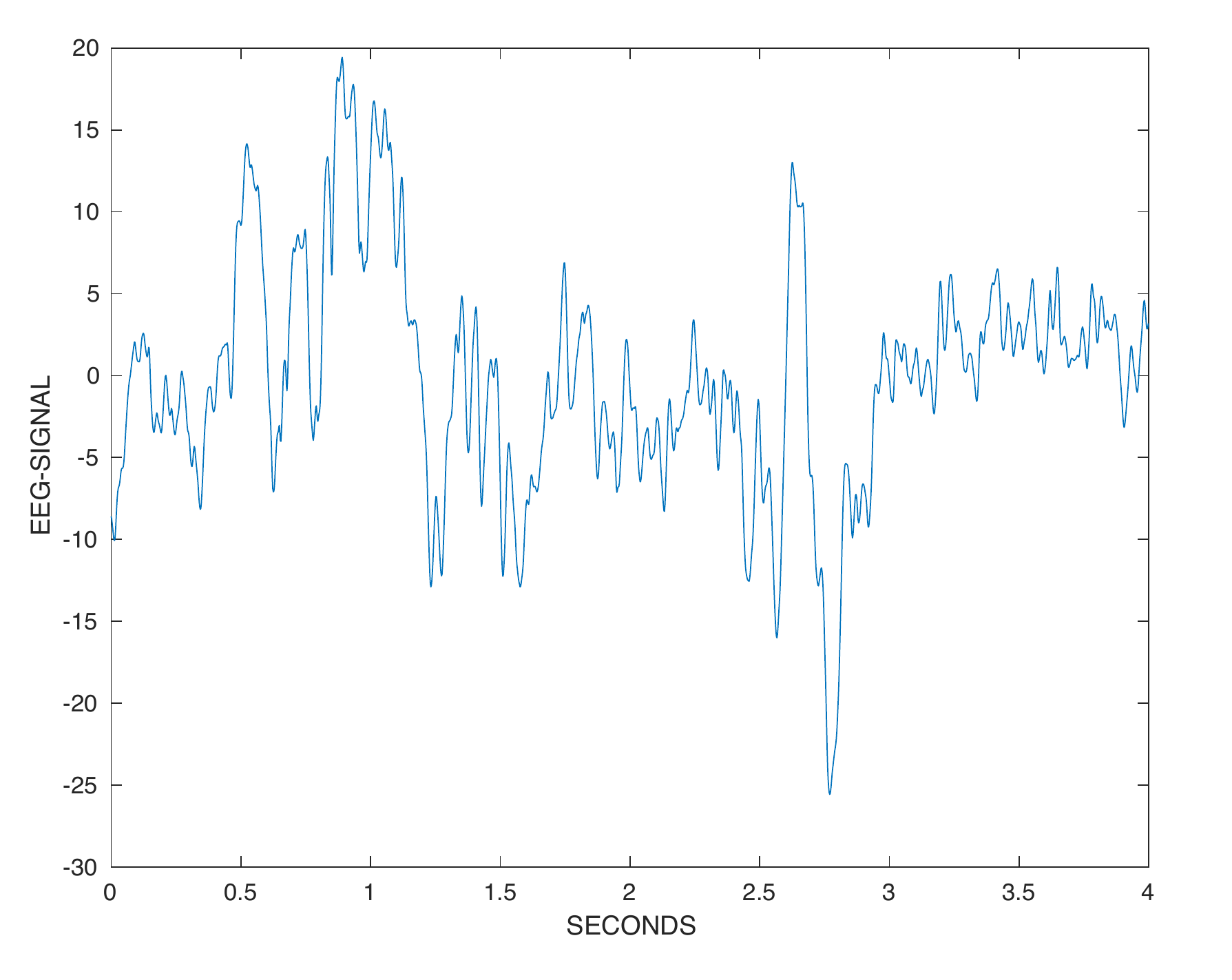}}
  \centerline{\footnotesize{$a)$ EEG Signal}}
\end{minipage}
\hspace{0.01\linewidth}
\begin{minipage}[b]{0.45\linewidth}
  \centering
  \centerline{\includegraphics[width=\linewidth]{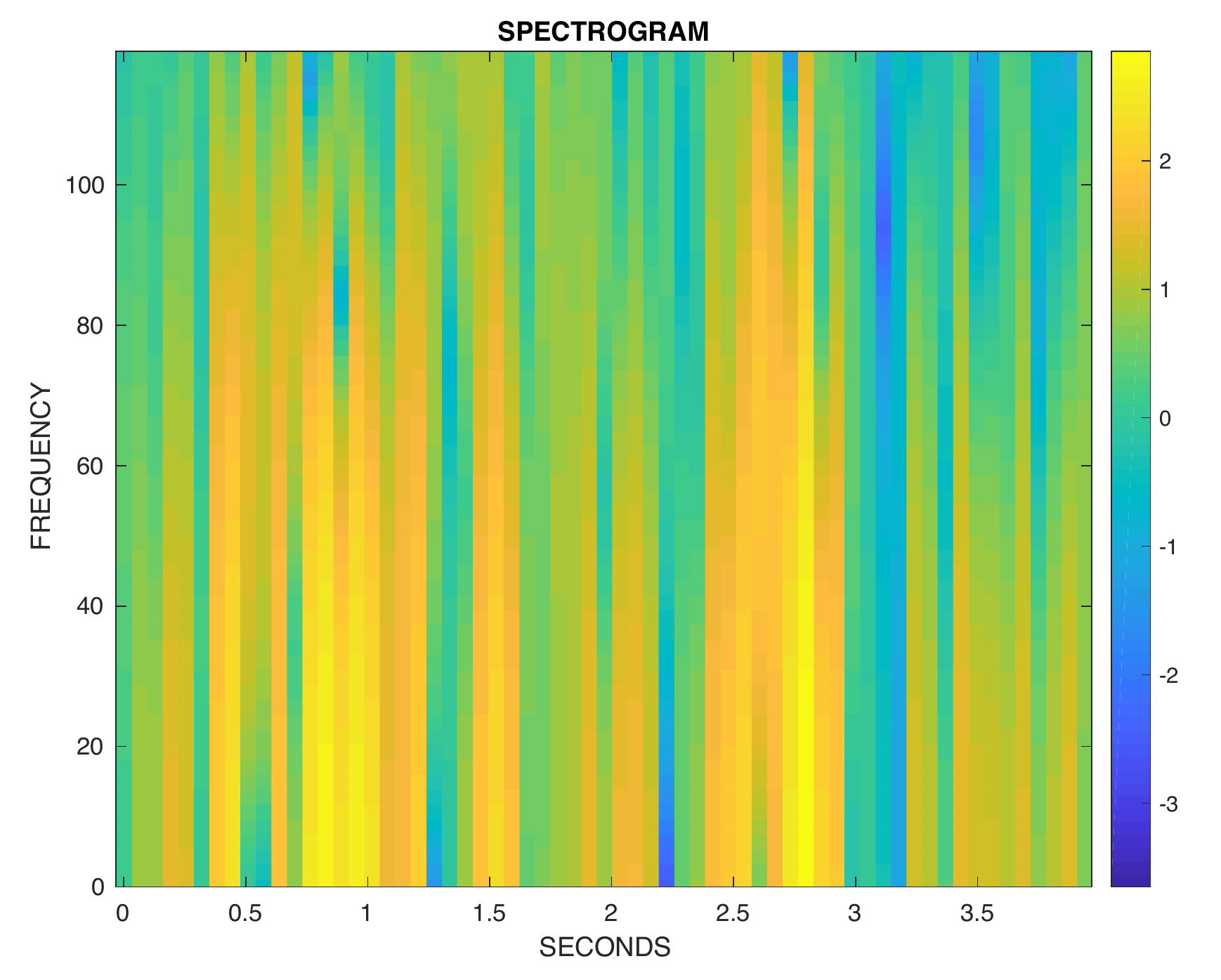}}
  \centerline{\footnotesize{$b)$ Log-Spectrogram}}
\end{minipage} 
\
\caption{EEG Log-spectrogram representation.}
\label{EEG_SP}
\end{figure}

In EEG signal analysis, it is often useful to characterize a signal by the time localization of its frequency components. This is particularly important for A-phase classification, since the A-phase sub-types are defined in terms of their spectral content. A popular time-frequency decomposition technique is the short-time Fourier transform, which has a good characterization in both domains. The magnitude of the short-time Fourier transform is also known as a spectrogram \cite{haghighi1997}. In the present work, we compute the spectrogram $S(x,y)$ representation of the A-phases and N-phases (4s segments), so that the x-axis represents time and the y-axis represents the frequency, and at each coordinate $S(x,y)$ represents the energy of a frequency component at a particular time. Under this representation, a signal can be seen as an image in such a way that it is possible to use deep convolutional neural networks designed for image classification. Furthermore, to emphasize differences between classes, a logarithm function is applied to the spectrograms, since the dynamic range of the amplitude is too large and masks the discriminatory information. Fig \ref{EEG_SP} shows a 4-second A-phase sample and its log-spectrogram; notice that the frequency content of the EEG signal ranges from 0.5 to 100 Hertz, thus we are using 120Hz as the maximum frequency for our analysis.  

\subsubsection{CNN-Architecture}\label{CNN_architecture}
To differentiate A-phases  and N-phase represented by their spectrograms, we used a Convolutional Neural Network (CNN). It is well known \cite{krizhevsky2012,arel2010,ciresan2012} that this kind of deep neural network has some properties which make them an efficient alternative for supervised classification. Its main attribute is to extract image features by means of learnable convolution operators, whose responses in turn are transmitted to a full neural network to carry out an accurate classification.

\begin{figure*}[t]
\centering
\begin{minipage}[b]{0.9\linewidth}
  \centering
  \centerline{\includegraphics[width=\linewidth]{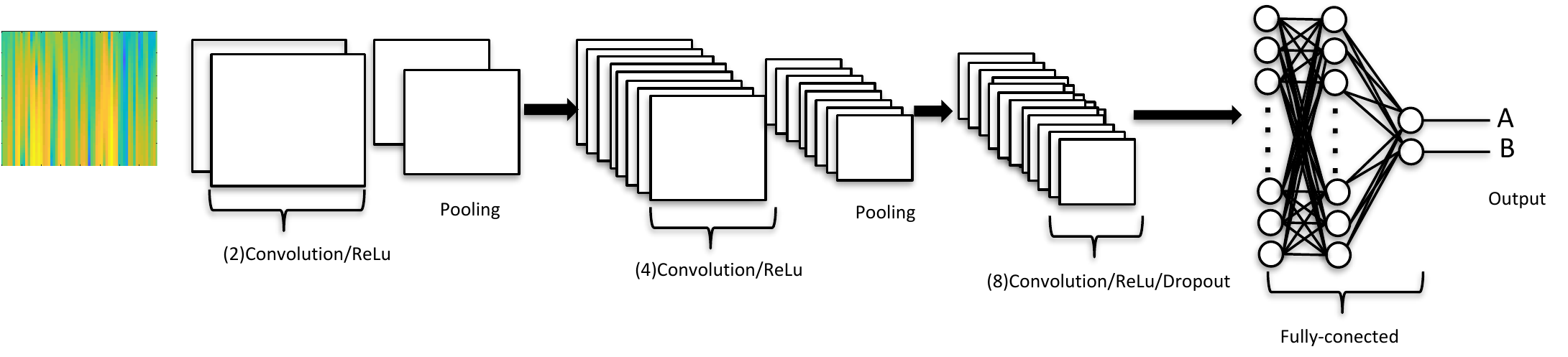}}
  \centerline{\footnotesize{$a)$ A/N Deep Classifier}}
\end{minipage}
\hspace{0.01\linewidth}
\begin{minipage}[b]{0.9\linewidth}
  \centering
  \centerline{\includegraphics[width=\linewidth]{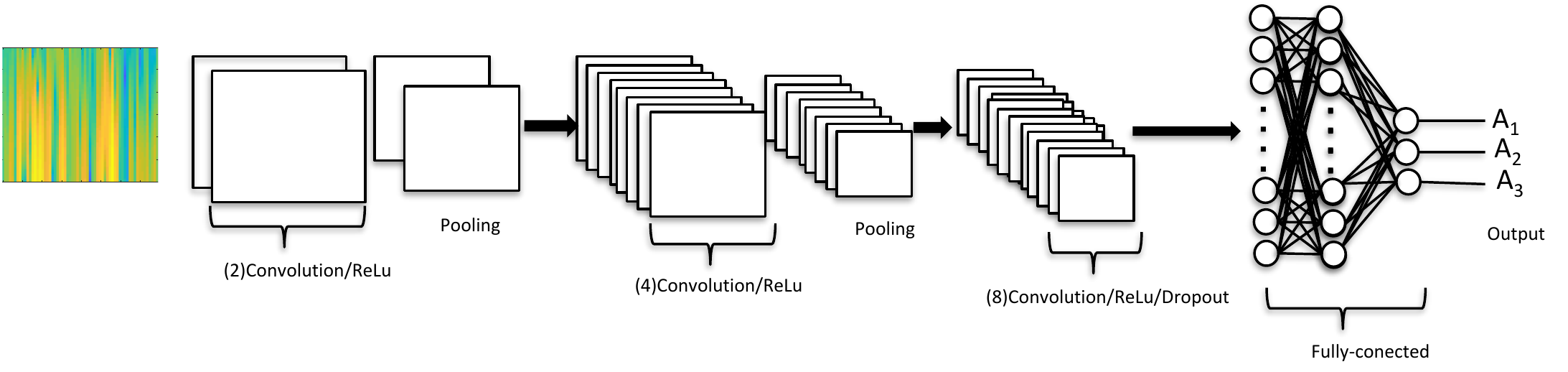}}
  \centerline{\footnotesize{$b)$ A1/A2/A3 Deep Classifier}}
\end{minipage} 
\
\caption{Proposed CNN architectures for A-phase classification.}
\label{CNN_architectures}
\end{figure*}

The CNN-architectures used in this work are depicted in Fig. \ref{CNN_architectures}. The first network, panel $a)$, is composed by a first convolution layer of two $3 \times 3$ kernels followed by a rectified linear unit (shortened to ReLu). The second layer is similar to the previous one except it uses four $3 \times 3 \times 2$ kernels.  A max-pooling operation is applied to each of the first two convolution+ReLu layers with the aim of reducing the size of the data by half.  The next layer has eight $3 \times 3 \times 4$ kernels, a ReLu, and a dropout method to randomly ignore $50\%$ of its outputs during training in order to reduce overfitting. Finally, a two full-connected neural network, which is a classical artificial network, is connected to a two output layer to classify A-phases and N-phases. A detailed description of each layer and their outputs can be found in Table \ref{tab:cnn}.

The second network, shown in panel $b)$, is nearly identical to the first one, but differs in the output layer, which in this case is designed to classify the A-phases in three classes  $A_1,A_2$, and $A_3$.

For the training of both CNN, the same set-up was used, which consisted in stochastic gradient descent with momentum equal to $0.9$, initial learning rate was set to $0.01$, the maximum number of epochs for training was $200$. Training was performed using mini-batches with 128 observations at each iteration, and a loss function based on cross entropy with $l_2$-regularization. It is worth mentioning that the final architecture as well as the parameter values were chosen based on experimentation with several alternatives.

\begin{table*}
\begin{tabular}{|lll|}
\hline
{\bf Layer} & {\bf Description} & {\bf Output-shape} \\ \hline
Input & (120,64,1) images &  \\
2D Convolution & 2 3x3 convolutions, stride [1  1] & (120,64,2) \\
Batch Normalization & Batch normalization & (120,64,2) \\
ReLu & ReLu & (120,64,2) \\
Max Pooling & 2x2 max pooling, stride [2  2]  & (60,32,2) \\
2D Convolution & 4 3x3x2 convolutions, stride [1  1] & (60,32,4) \\
Batch Normalization & Batch normalization & (60,32,4) \\
ReLu & ReLu & (60,32,4) \\
Max Pooling & 2x2 max pooling, stride [2  2]  & (30,16,4) \\
2D Convolution & 8 3x3x4 convolutions, stride [1  1]  & (30,16,8) \\
Batch Normalization & Batch normalization & (30,16,8) \\
ReLu & ReLu & (30,16,8) \\
Dropout & 50\% dropout & (30,16,8) \\
Fully Connected & Dense layer & 3840 \\
Fully Connected & Dense layer & 3840 \\
Output & Classification Output & 2 units (A,N) \\
& & or 3 units (A1,A2,A3) \\ \hline
\end{tabular}
\caption{Description of the layers of the proposed deep networks.}
\label{tab:cnn}
\end{table*}

Once the architecture was defined, it is important to prepare the data for training. As described above, two networks are trained for each subject using the log-spectrograms corresponding to the A-phase and N-phase segments. The first network is designed to distinguish between A-phases and N-phases; therefore, it will be trained using both kinds of segments (A and N). The second network is designed to determine the sub-type of a given A-phase; thus it will be trained using A-phases only. In order to avoid training bias due to the differences in the number of data samples in each class, the training data was balanced such that all classes had the same number of elements. To do this, the training set for classes with less data was augmented by copying randomly (with replacement) from the same set until all classes were equally represented.

\section{Results and discussion}

In order to test the proposed methodology, we have trained and evaluated multiple ad hoc CNNs for each of the nine subjects from the Physionet database. One of the main questions that arise from this study is how many training data are required to obtain a competitive classifier, with respect to the accuracy rates achieved by previous studies (i.e., between 70\% and 80\%). To answer this question, we have trained the networks using different percentages of each subject's data, and evaluated the network with the remaining data. These percentages are 12.5\%, 25\%, 37.5\% and 50\%. Assuming that the sleep time for each subject is approximately 8 hours, then the percentages of training data roughly correspond to 1, 2, 3 and 4 hours of sleep time. However, training samples were chosen randomly for each subject from all of the data, not specifically from the first hours of sleep.

\subsection{Accuracy vs percentage of training data}

Two experiments were performed. In the first one, which will denoted as A/N, the goal is to distinguish between A-phases (regardless of their sub-type) and N-phases (non A-phases). In the second experiment, denoted as A1/A2/A3, the goal is to classify A-phases according to their sub-type (A1, A2 or A3); in this case, N-phases are not used either for training or testing.

A total of 20 networks with the proposed architecture were trained for each experiment, for each subject and each percentage of training data. For each network, the training set was randomly chosen  from the corresponding subject's data, while the remaining data (from the same subject) was used for evaluating the network's performance. Each training set is composed of a different number of samples from each class; however, training batches were chosen from the training data by sampling with replacement, so that the same number of samples from each class was used in each batch, avoiding biasing the network towards the most probable class. Once a network was trained, its classification accuracy was measured as the percentage of test data points that were correctly classified. For each experiment, subject and training percentage, the average accuracy and its standard deviation were computed from the 20 runs.

\begin{figure*}
\centering
\begin{tabular}{p{5.5cm}p{5.5cm}}
\resizebox{5.5cm}{!}{\includegraphics{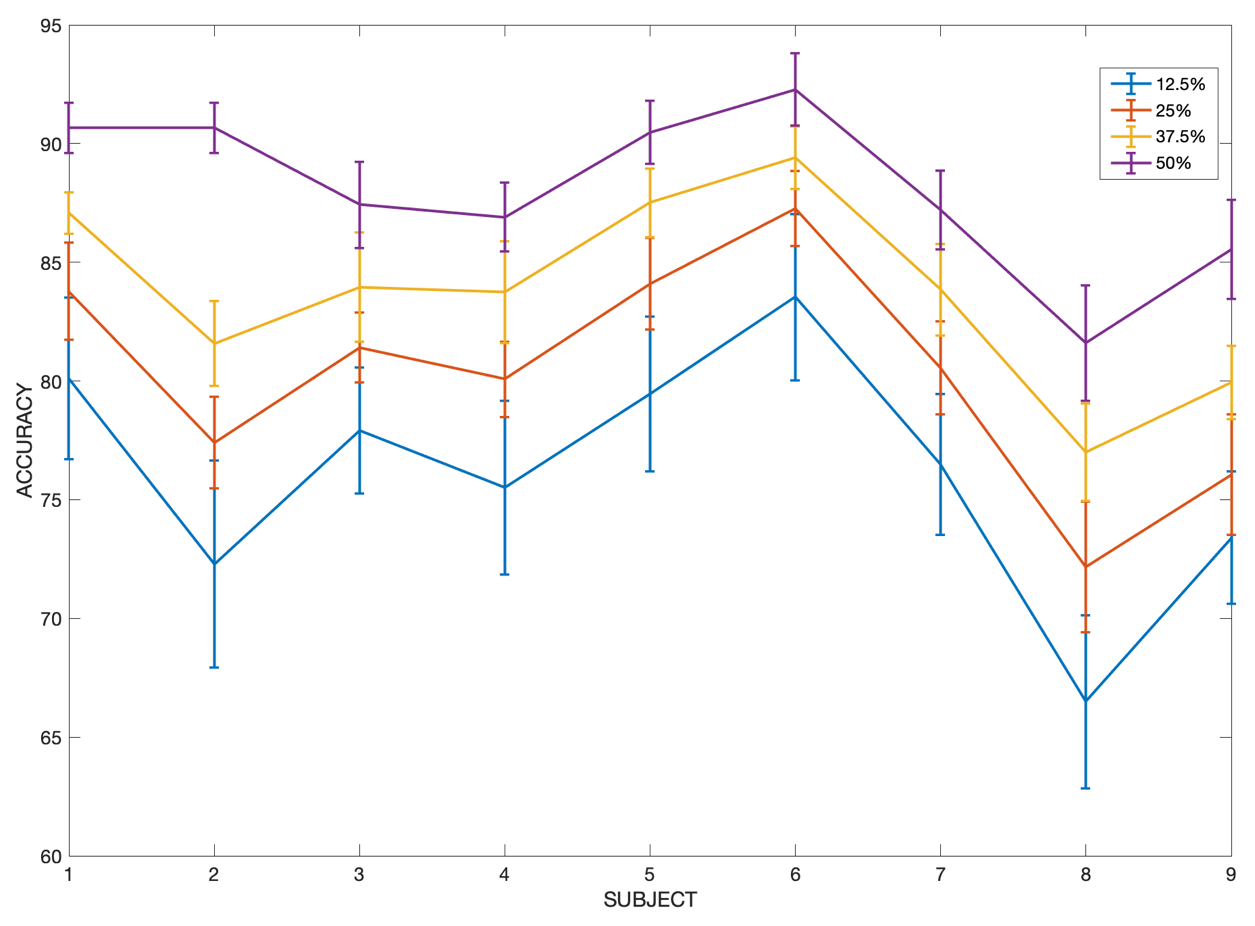}} & 
\resizebox{5.5cm}{!}{\includegraphics{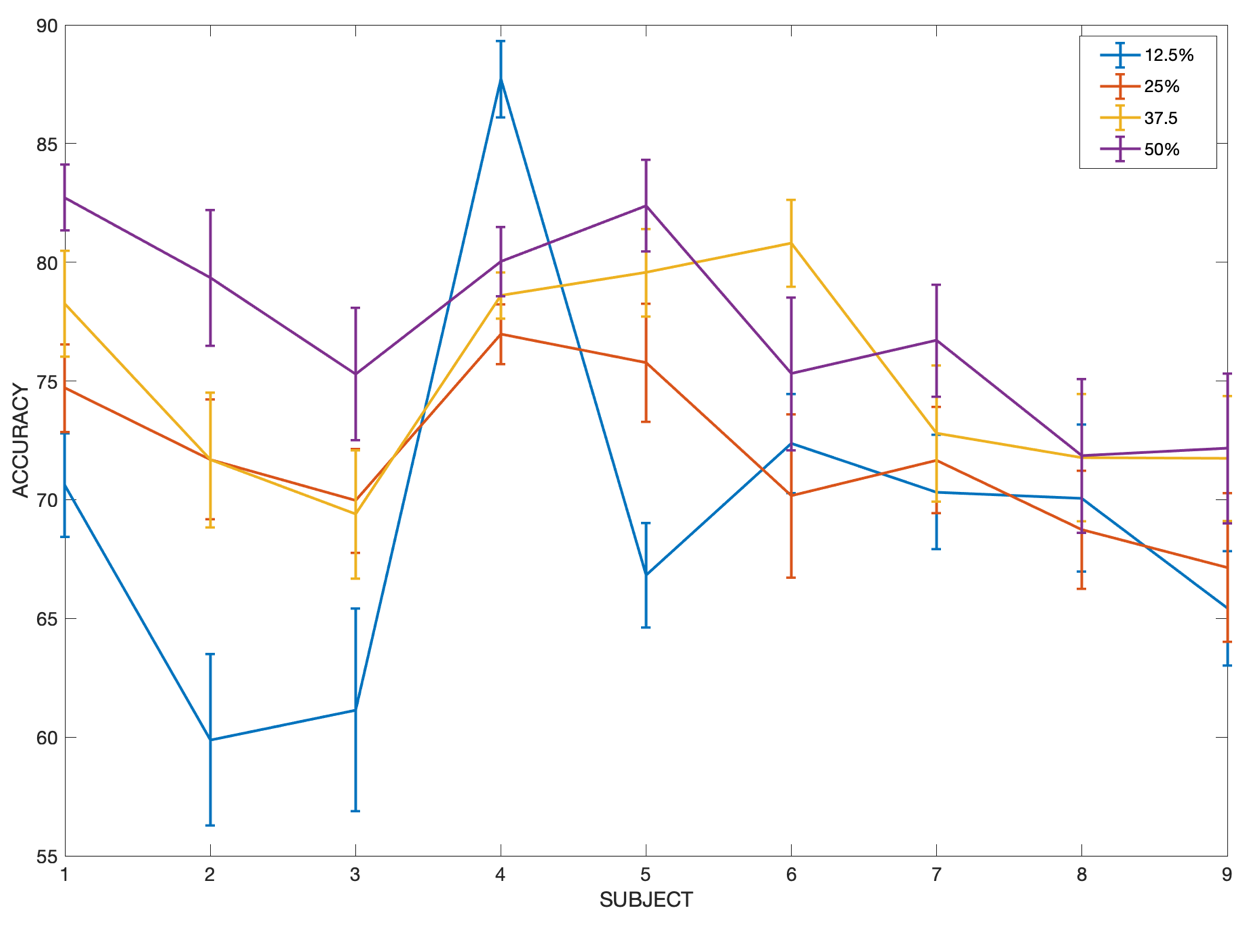}} \\
\begin{center}(a) A/N\end{center} & \begin{center}(b) A1/A2/A3\end{center}
\end{tabular}
\caption{Average accuracy of 20 convolutional networks trained to classify A-phases for each subject, using different percentages of training data (according to the legend). The whiskers in each graph represent the standard deviation. (a) Networks trained to distinguish between A-phases (regarless of their sub-type) and N-phases. (b) Networks trained to classify A-phases according to their sub-type. }
\label{fig:training_percentage}
\end{figure*}

Results for both experiments are shown in Figure \ref{fig:training_percentage}, whereas the average accuracy across the nine subjects is shown in Table \ref{tab:results_training}. For A/N classification, it is clear from Figure \ref{fig:training_percentage}a that even the smallest training percentage (12.5\%) is enough to obtain competitive results (accuracy $>$ 70\%) for most subjects, with the notable exception being Subject 8. As expected, the accuracy increases as the percentage of training data is increased, reaching accuracies over 90\% for some subjects. It is worth noting that 12.5\% of each subject's data represents a very small number of EEG segments, which averages 50 samples per class, according to the lower section of Table \ref{tab:results_training}; in contrast to previous works where a large number of data segments from many subjects are required to train the classifier (not to mention the need to compute a large number of features for each EEG segment). Because of this, training an ad-hoc network for a given subject is a relatively quick process. In these experiments, the average training time per subject lies between 15 and 40 seconds with a Matlab implementation running under a single CPU thread, and depends on the number of training samples. With the aid of a modern GPU, training times can be even smaller.

On the other hand, results from A1/A2/A3 classification (Figure \ref{fig:training_percentage}b) do not follow such a regular pattern; in general, they show lower scores and large variations from one subject to another. This is due to a number of reasons: first, multi-class classification is usually a harder problem than binary classification and often requires larger amounts of training data; however, since only A-phases are used to train these classifiers, we are using approximately half the number of training samples that were used for the A/N classifiers. For instance, in the case of the A/N classifiers, one can obtain a competitive score (\~ 80\%) using approximately 100 training samples per class (around 25\% of the data, see Table \ref{tab:results_training}). One could think that a similar amount of training samples per class would yield a similar score for the sub-type classifier; however, this means using a total of 300 training samples, which is approximately a 75\% of the average number of A-phases which occur during a whole-night's sleep. In other words, for a given percentage of training data, the number of training data per class in the A1/A2/A3 classifier is roughly one third of the corresponding number for the A/N classifier. Moreover, the distribution of A1, A2 and A3 phases is quite unbalanced; for a given subject, there may not enough A2 and A3 phases to meet the required 100 for each class. From these reasons, it should be clear why the sub-type classification is considerably harder, but in average, still benefits from a larger number of training samples, as the results in Table \ref{tab:results_training} demonstrate.

\begin{table*}
\centering
{\scriptsize
\begin{tabular}{|c|cccc|}
\hline
\multicolumn{5}{|c|}{Average accuracy}  \\ \hline
{\bf Classes} & {\bf 12.5\%} & {\bf 25\%} & {\bf 37.5\%} & {\bf 50\%} \\ \hline
{\bf A / N} & 76.13 & 80.31 & 83.78 & 88.09 \\ \hline
{\bf A1/A2/A3} & 69.37 & 71.87 & 74.95 & 77.31 \\ \hline
\hline
\multicolumn{5}{|c|}{Number of training/testing samples (A-phases only)}  \\ \hline
{\bf Subject} & {\bf 12.5\%} & {\bf 25\%} & {\bf 37.5\%} & {\bf 50\%} \\ \hline
1 & (45,11,10)/471  & (90,23,20)/404  & (136,35,30)/336  & (181,47,40)/269 \\
2 & (23,9,11)/309  & (46,18,23)/256  & (69,27,35)/221  & (93,36,47)/176 \\
3 & (17,13,13)/312  & (35,26,27)/267  & (52,39,40)/224  & (70,53,54)/178 \\
4 & (57,3,7)/479  & (115,6,15)/410  & (173,9,22)/342  & (231,12,30)/273 \\
5 & (37,14,11)/450  & (75,28,23)/386  & (113,43,35)/321  & (151,57,47)/257 \\
6 & (38,12,5)/393  & (76,24,10)/338  & (115,37,15)/281  & (153,49,21)/225 \\
7 & (26,7,6)/277  & (53,14,12)/237  & (79,21,18)/198  & (106,28,24)/158 \\
8 & (20,4,7)/228  & (41,8,15)/195  & (61,12,22)/164  & (82,17,30)/130 \\
9 & (29,10,6)/320  & (58,20,12)/275  & (88,30,18)/229  & (117,40,25)/183 \\ \hline
Mean & 50.11/359.88 & 101.44/307.55 & 153/257.33 & 204.55/204.44 \\ \hline
\end{tabular}}
\caption{Upper section: Average accuracy across all subjects for CNNs trained to classify A-phases with different percentages of training data. Lower section: number of A-phases used for training (A1, A2, A3) and testing, per subject. The last row shows the average number of training/testing samples across all subjects. The number of N-phases used for training the A/N classifiers are approximately equal to the total number of A-phases for each subject.}
\label{tab:results_training}
\end{table*}

\subsection{Retraining with expert-validated data}

Results from the previous section suggest that the proposed deep network architecture is capable of achieving a competitive accuracy ($>$ 70\%) with as little as 100 training samples among all classes. This is roughly equivalent to annotating 1 hour of EEG data for A/B classification, or two hours of data for A1/A2/A3 classification, from a polysomnographic recording of 8 hours. On the other hand, accuracy percentages will, in average, increase as the expert annotates more segments, but this of course requires additional work from the expert.

One way to aid the expert in this task, is to train the network using a small set of training samples (e.g., $\sim 50$ segments), and then use that network to classify the rest of the segments. Then, the expert could validate some of the classified segments, for instance, those segments for which the expert has a high certainty that they were correctly classified. This validation process should require less involvement from the expert than the annotated segments used for the initial training, since the user must only select some of the segments in which he or she agrees with the automatic classification. Once this has been done, the network can be re-trained with an extended data set composed of the original training set plus the expert-validated samples.

In order to test this idea, another set of experiments were performed where, for each subject, a network was trained using only 12.5\% of the subject's data, and then used to classify the remaining 87.5\% of the segments. Expert-validation was simulated by choosing a certain percentage (between 20\% and 50\%) of the data that had been correctly classified by the network. Then, the network was re-trained using both the original training data and the expert-validated data, and re-evaluated using the remaining data.

\begin{figure*}
\centering
\begin{tabular}{p{5.5cm}p{5.5cm}}
\resizebox{5.5cm}{!}{\includegraphics{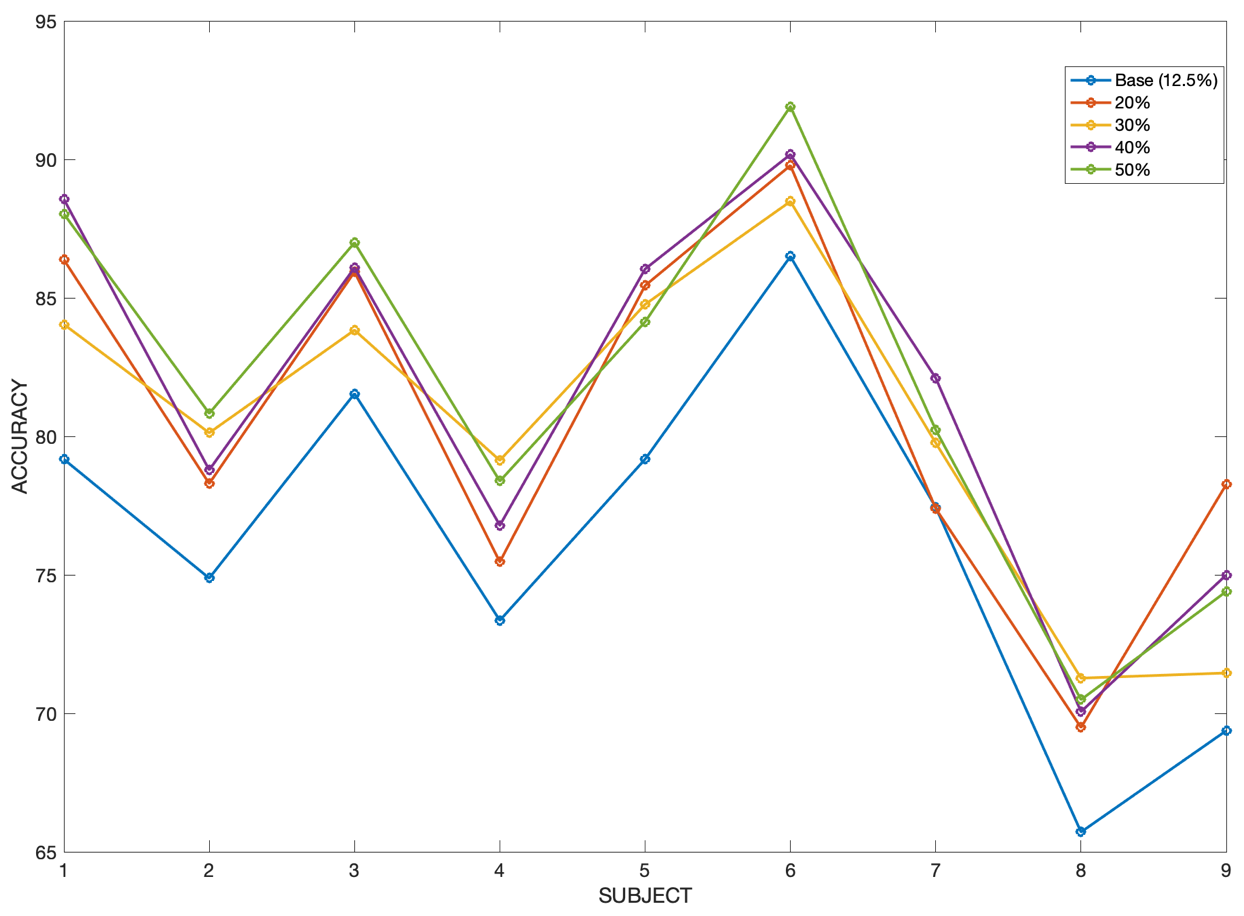}} & 
\resizebox{5.5cm}{!}{\includegraphics{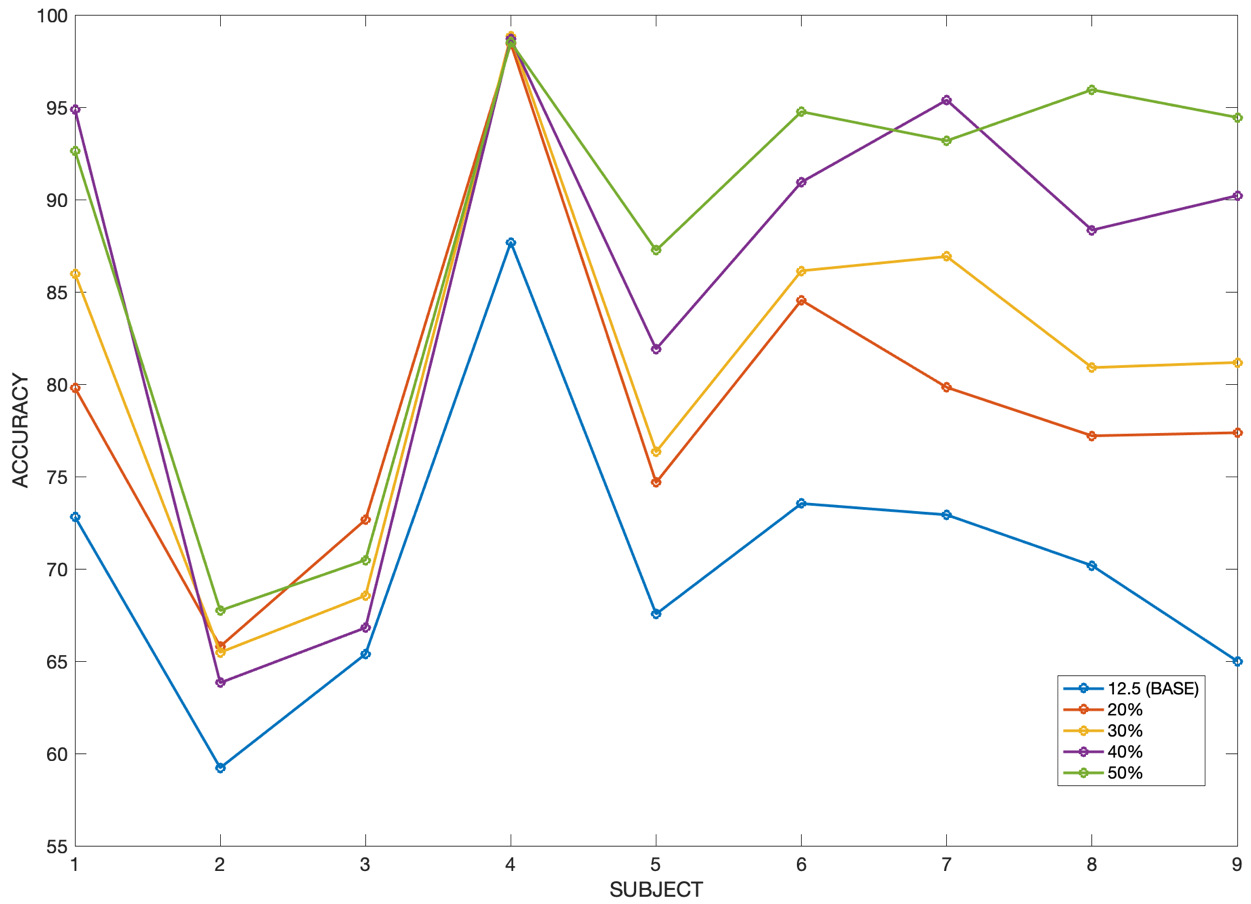}} \\
\begin{center}(a) A/N\end{center} & \begin{center}(b) A1/A2/A3\end{center}
\end{tabular}
\caption{Average accuracy of 20 convolutional networks trained to classify A-phases for each subject, using re-training with different percentages of expert-validated data (according to the legend). The base case corresponds to an initial training with 12.5\% of the available data for each subject. The whiskers in each graph represent the standard deviation. (a) Networks trained to distinguish between A-phases (regarless of their sub-type) and N-phases. (b) Networks trained to classify A-phases according to their sub-type. }
\label{fig:retraining_percentage}
\end{figure*}

Results from these experiments are shown in Figure \ref{fig:retraining_percentage} and summarized in Table \ref{tab:results_retraining}. In these results, the base case corresponds to a single training stage using 12.5\% of the available data for each subject (50 training samples, in average), while percentages from 20\% to 50\% represent the amount of expert-validated data that was used to augment the training set for a second training stage.

For the A/N classifier, Figure \ref{fig:retraining_percentage}a shows that even with as little as 20\% of expert-validated data, one obtains a significant increase in accuracy. However, using more than 20\% does not seem to yield significant benefits as the average accuracy stalls at around 81\%. In this case, it seems to be more beneficial to simply annotate more segments for the first training stage.

For sub-type classification, however, the results from re-training are very interesting. On one hand, the accuracy increases for most subjects as the re-training percentage increases. In all cases, re-training with at least 20\% of expert-validated data is highly beneficial. With only 20\% of expert-validated data, the proposed CNN achieves an average accuracy of nearly 79\%, and the accuracy increases up to 88.32\% with 50\% of user-validated data. This is very encouraging since A-phase sub-type classification has been shown to be a difficult problem, where previous works have reported accuracy values between 67\% and 71\% \cite{machado2016,machado2018,mostafa2018}. On the other hand, it is clear that the benefits of re-training with more than 20\% of the data depend on each subject, particularly in the cases of subjects for whom the number of A2- and A3- phases is considerably less or more than half the number of A1-phases, as with Subjects 2, 3 and 4 in this study.

\begin{table}
\centering
{\scriptsize
\begin{tabular}{|c|ccccc|}
\hline 
{\bf Classes} & {\bf Base accuracy} & {\bf 20\%} & {\bf 30\%} & {\bf 40\%} & {\bf 50\%} \\ \hline
{\bf A / N} & 76.34 & 80.71 & 80.31 & 81.51 & 81.71  \\ \hline
{\bf A1/A2/A3} & 70.47 & 78.92 & 81.14 & 85.66 & 88.32 \\ \hline
\end{tabular}}
\caption{Average accuracy across all subjects for CNNs trained to classify A-phases using re-training with different percentages of expert-validated data. The base case corresponds to an initial training with 12.5\% of the available data for each subject.}
\label{tab:results_retraining}
\end{table}

\section{Conclusions}
Two convolutional neural networks (CNNs) for the classification of EEG segments as A-phases or non A-phases, and the classification of A-phases according to their sub-type (A1 / A2 / A3), have been introduced in this work. There are two important differences in the proposed method with respect to other methods proposed in the literature: first, instead of computing a number of static features for each EEG segment, we compute the log-spectrogram of the signal and treat it as the input image for the CNNs. Second, we propose to train an ad-hoc classifier for each subject; this approach requires some involvement from an expert but achieves more accurate results in only a fraction of the time required for the fully manual annotation.

The main findings are: (1) Semi-automatic A-phase annotation with assistance from an expert, using ad-hoc classifiers trained for each subject could provide a good alternative to fully automatic classifiers whose accuracy is limited. (2) The proposed approach is capable of producing competitive results when there is little involvement from the expert, and very good results with additional expert validation. (3) Results vary from subject to subject, but as a rule of thumb, results are consistent for those subjects for whom the ratio of the number of A2- and A3- phases with respect to the number of A1-phases is close to 0.5.

It is important to recall that, in this work, it is assumed that the A-phase onset times are known beforehand. However, according to \cite{mendez2014,mendez2016}, A-phase onsets are relatively easy to detect, either manually or automatically. The A/N classifier could be then applied to filter out false positives, and then the A1/A2/A3 classifier can be used to detect the A-phase sub-type, based on the first 4s of the A-phase. With these ideas, we are currently working towards a more efficient semi-automatic A-phase annotation system.

\bigskip

\bibliographystyle{unsrt}

\bibliography{cap_cnn}

\end{document}